\documentclass[letterpaper,twocolumn,english,prl,superscriptaddress,showpacs,longbibliography,fixfloat,notitlepage]{revtex4-1}
\usepackage[latin9]{inputenc}
\usepackage{xcolor}
\usepackage{pdfcolmk}
\usepackage{amsmath}
\usepackage{amssymb}
\usepackage{graphicx}
\PassOptionsToPackage{normalem}{ulem}
\usepackage{ulem}

\makeatletter

\pdfpageheight\paperheight
\pdfpagewidth\paperwidth

\providecolor{lyxadded}{rgb}{0,0,1}
\providecolor{lyxdeleted}{rgb}{1,0,0}
\DeclareRobustCommand{\lyxsout}[1]{\ifx\\#1\else\sout{#1}\fi}

\usepackage[colorlinks,citecolor=darkblue,linkcolor=darkred,urlcolor=darkblue] {hyperref}
\usepackage{xcolor}
\definecolor{darkblue}{rgb}{0.1,0.2,0.6} 
\definecolor{darkred}{rgb}{0.8,0.1,0.2}
\renewcommand{\BibitemShut}[1]{}

\makeatother

\usepackage{babel}
\begin{document}
\global\long\def\E{\mathrm{e}}
\global\long\def\D{\mathrm{d}}
\global\long\def\I{\mathrm{i}}
\global\long\def\mat#1{\mathsf{#1}}
\global\long\def\vec#1{\mathsf{#1}}
\global\long\def\cf{\textit{cf.}}
\global\long\def\ie{\textit{i.e.}}
\global\long\def\eg{\textit{e.g.}}
\global\long\def\vs{\textit{vs.}}
 \global\long\def\ket#1{\left|#1\right\rangle }

\global\long\def\etal{\textit{et al.}}
\global\long\def\tr{\text{Tr}\,}
 \global\long\def\im{\text{Im}\,}
 \global\long\def\re{\text{Re}\,}
 \global\long\def\bra#1{\left\langle #1\right|}
 \global\long\def\braket#1#2{\left.\left\langle #1\right|#2\right\rangle }
 \global\long\def\obracket#1#2#3{\left\langle #1\right|#2\left|#3\right\rangle }
 \global\long\def\proj#1#2{\left.\left.\left|#1\right\rangle \right\langle #2\right|}

\title{Emergent locality in systems with power-law interactions}

\author{David J. Luitz}
\email{dluitz@pks.mpg.de}

\affiliation{Max-Planck-Institut für Physik komplexer Systeme, Nöthnitzer Str. 38,
01187 Dresden, Germany}

\affiliation{Department of Physics, T42, Technische Universität München, D-85748
Garching, Germany}

\author{Yevgeny Bar Lev}
\email{yevgeny.barlev@weizmann.ac.il}

\affiliation{Department of Condensed Matter Physics, Weizmann Institute of Science,
Rehovot 76100, Israel}

\affiliation{Max-Planck-Institut für Physik komplexer Systeme, Nöthnitzer Str. 38,
01187 Dresden, Germany}
\begin{abstract}
Locality imposes stringent constraints on the spreading of information
in nonrelativistic quantum systems, which is reminiscent of a ``light-cone,''
a causal structure arising in their relativistic counterparts. Long-range
interactions can potentially soften such constraints, allowing almost
instantaneous long jumps of particles, thus defying causality. Since
interactions decaying as a power-law with distance, $r^{-\alpha}$,
are ubiquitous in nature, it is pertinent to understand what is the
fate of causality and information spreading in such systems. Using
a numerically exact technique we address these questions by studying
the out-of-time-order correlation function of a representative generic
system in one-dimension. We show that while the interactions are long-range,
their effect on information spreading is asymptotically negligible
as long as $\alpha>1$. In this range we find a complex compound behavior,
where after a short transient a fully local behavior emerges, yielding
asymptotic ``light-cones'' virtually indistinguishable from ``light-cones''
in corresponding local models. The long-range nature of the interaction
is only expressed in the power-law leaking of information from the
``light-cone,'' with the same exponent as the exponent of the interaction,
$\alpha$. Our results directly imply that all previously obtained
rigorous bounds on information spreading in long-range interacting
systems are not tight, and thus could be improved.
\end{abstract}
\maketitle
\emph{Introduction.\textemdash }Special relativity prohibits passing
signals faster than the speed of light, embodying the concept of causality.
All causal information is hence strictly contained within a light-cone.
In contrast, the speed of light is not directly relevant for \emph{nonrelativistic}
systems. Nevertheless, it was rigorously shown by Leib and Robinson
that remnants of causality exist also in nonrelativistic quantum systems
with short-range interactions \cite{Lieb1972}. While \emph{most}
of the causal information travels within a ``light-cone,'' some
of it ``leaks'' outside with tails exponentially decaying with the
distance from the ``information front.''. The shape of this ``light-cone''
can be obtained from
\begin{equation}
C_{x}\left(t\right)=\left\Vert \left[\hat{A}_{i}\left(t\right),\hat{B}_{i+x}\right]\right\Vert ,\label{eq:otoc}
\end{equation}
where $\hat{A}_{i}\left(t\right)$ and $\hat{B}_{i+x}$ are local
Hermitian operators written in the Heisenberg picture operating on
sites $i$ and $i+x$ and $\left\Vert .\right\Vert $ is a norm in
the operator space. Lieb and Robinson proved that for short-range
interacting Hamiltonians $C_{x}\left(t\right)\leq\exp\left[\lambda\left(t-x/v\right)\right]$,
where $\lambda$ is a constant and $v$ is the Lieb-Robinson (LR)
velocity, which depends on the microscopic properties of the model
\cite{Bohrdt2016}. It is important to note, that since the LR bound
is a bound on an \emph{operator}, it is\emph{ independent} of the
initial state of the system. If the norm is chosen to be the normalized
Frobenius norm $\left[\left\Vert \hat{A}\right\Vert ^{2}=\mathcal{N}^{-1}\text{Tr}\left(\hat{A}^{\dagger}\hat{A}\right)\right]$,
where $\mathcal{N}$ is the Hilbert space dimension, $C_{x}\left(t\right)$
is directly related to the out-of-time-order correlation function
(OTOC), which was first introduced by Larkin and Ovchinnikov \cite{Larkin1969}.
Since in this work we only use the Frobenius norm, we will use $C_{x}\left(t\right)$
and OTOC interchangeably. Larkin and Ovchinnikov noted that for quantum
systems with a semiclassical limit the OTOC embodies a signature of
classical chaos in the corresponding quantum system. 
\begin{figure}[th]
\includegraphics[width=1\columnwidth]{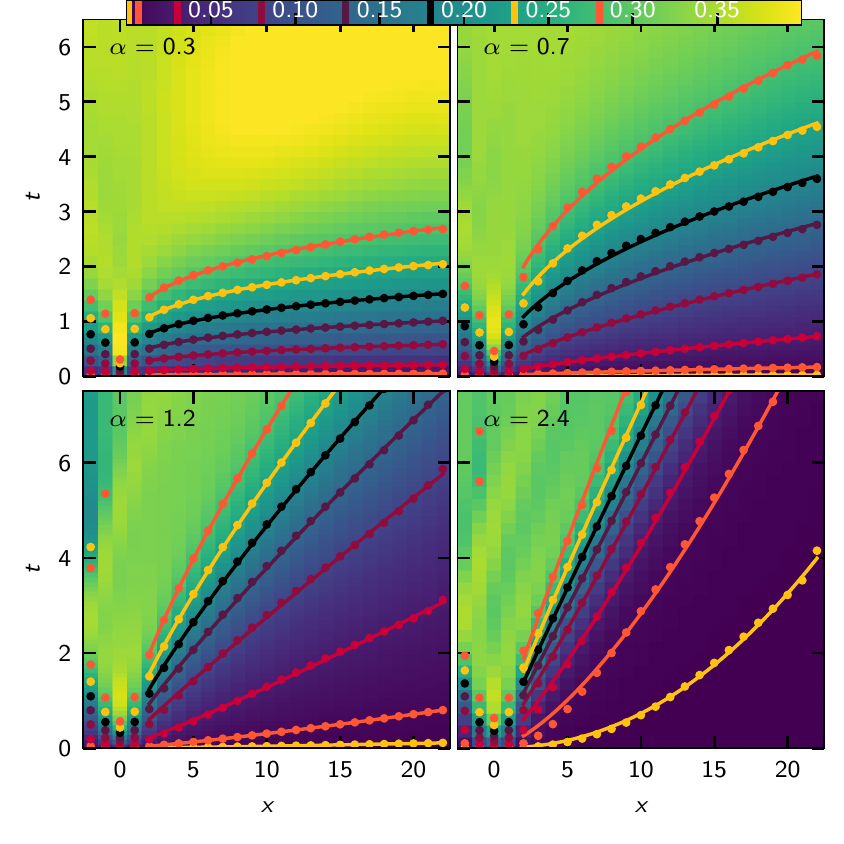}\caption{\label{fig:overview}Spreading of the out-of-time-order correlator
(OTOC), $C_{x}\left(t\right)$ for various interaction exponents $\alpha$.
The points correspond to discrete contour lines of the OTOC, calculated
by locating where $C_{x}\left(t\right)$ exceeds certain thresholds
$\theta$, which are indicated in the color bar. Full lines are power-law
fits to these contour lines. Here, we use open boundary conditions
and $L=25$ and take $i=3$. }
\end{figure}
In the semiclassical limit the commutator is replaced by Poisson brackets
and the choice of the operators $\hat{A}\left(t\right)\to q\left(t\right)$
and $\hat{B}\to p$, gives $C\left(t\right)\sim\hbar^{2}\left(\partial q\left(t\right)/\partial q\right)^{2}$.
The OTOC therefore measures the sensitivity of classical trajectories
to their initial conditions, which for chaotic systems implies that
it grows exponentially in time, $C\left(t\right)\sim\exp\left[2\lambda_{L}t\right]$,
where $\lambda_{L}$ is the classical Lyapunov exponent. For quantum
systems with a \emph{finite} local Hilbert space dimension the OTOC
is bounded from above and its growth saturates, indicating a complete
loss of local information \cite{Bohrdt2016,Luitz2017}. Numerically
exact simulations show that such systems do \emph{not} exhibit a finite
regime of exponential growth \cite{Luitz2017}. 

For a large number of physical systems, with power-law decaying interactions,
$r^{-\alpha}$, the LR bound doesn't hold. Such systems include conventional
condensed matter systems such as nuclear spins \cite{Alvarez2015a},
dipole-dipole interactions of vibrational modes \cite{Levitov1989,Levitov1990,Aleiner2011},
Frenkel excitons \cite{Agranovich2008}, nitrogen vacancy centers
in diamond \cite{Childress2006,Balasubramanian2009,Neumann2010,Weber2010,Dolde2013}
and polarons \cite{Alexandrov1996}, but also molecular and atomic
systems where interaction can be dipolar \cite{Saffman2010,Aikawa2012,Lu2012,Yan2013,Gunter2013,DePaz2013},
van der Waals like \cite{Saffman2010,Schauss2012}, or even of variable
range \cite{Britton2012,Islam2013,Richerme2014a,Jurcevic2014a}. While
the LR bound doesn't hold, naively one can expect an enhancement of
the ``causal'' region, and faster than ballistic spreading of information.
Indeed, for $\alpha>d$ (where $d$ is the dimension of the system)
the LR bound was generalized by Hastings and Koma, who showed that
the causal region becomes at most logarithmic, $t\sim\log x$ \cite{Hastings2006}.
This result was subsequently improved to an algebraic ``light-cone,''
$t\sim r^{\xi}$ for $\alpha>2d$ and $0<\delta<1$ \cite{Foss-Feig2015}.
A LR-type bound was also obtained for $\alpha<d$ after a proper rescaling
of time \cite{Storch2015,Kastner2017}. It is currently an open question
how tight LR-type bounds are for\emph{ long-range} interacting systems,
as also, how universal the spreading of information \emph{is} in these
systems. Interestingly this question was \emph{not} addressed directly
by considering $C_{x}\left(t\right).$ Instead, previous studies considered
spatial one-time correlation functions, $K_{x}\left(t\right)\equiv\tr\left(\hat{\rho}_{0}\left(t\right)\hat{A}_{i}\hat{B}_{i+x}\right)$,
where $\hat{\rho}_{0}$ was taken to be either some special state
\cite{Eisert2013,Gong2014,Buyskikh2016} or a state resulting from
a quench from the groundstate \cite{Hauke2013,Cevolani2015,Cevolani2016,Buyskikh2016,Maghrebi2016,VanRegemortel2016,Lepori2017,Cevolani2017,Frerot2018}
(c.f. Ref.~\cite{Chen2017a} for a recent study of the OTOC). Unlike
$C_{x}\left(t\right)$ the correlation functions $K_{x}\left(t\right)$
are experimentally measurable, but depend on the initial state \cite{Eisert2013}
as also the microscopic details of the model \cite{Cevolani2015,Cevolani2016,Buyskikh2016,Frerot2018},
and therefore do \emph{not} have direct implications on the tightness
of LR-type bounds (unless a supremum over \emph{all} initial states
is taken). 

The spreading of correlations $K_{x}\left(t\right)$ was studied analytically
in long-range Kitaev chains, which are solvable quadratic models \cite{Storch2015,Cevolani2015,Buyskikh2016,VanRegemortel2016,Cevolani2016,Lepori2017,Frerot2018}.
It was shown that $K_{x}\left(t\right)$ correlations show linear
``light-cones'' for $\alpha>d+1$, \cite{Cevolani2016}, and ``light-cones''
with \emph{suppressed} causal region for $d<\alpha<d+1$. These results
suggest (but don't imply!) that all known LR-type bounds are not tight.
For $\alpha<d$, where no LR-type bounds apply, these systems show
instantaneous (in the thermodynamic limit) spreading of correlations
\cite{Storch2015,Cevolani2015,Buyskikh2016,VanRegemortel2016,Cevolani2016,Lepori2017,Frerot2018},
unless time is rescaled with the system size \cite{Kastner2011,Bachelard2013,VanDenWorm2013,Storch2015,Kastner2017}.
One question which naturally arises, is how universal are results
obtained for integrable long-range models, which are special by definition?
To answer this question, one must consider \emph{generic} long-range
systems. Spreading of $K_{x}\left(t\right)$ correlations in such
systems were studied using numerically exact methods \cite{Hauke2013,Eisert2013,Schachenmayer2014,Buyskikh2016},
variational methods \cite{Cevolani2015} and by \emph{approximately}
reducing them to quadratic effective models, either by studying the
corresponding quasiparticle descriptions \cite{Cevolani2015,Cevolani2016,Cevolani2017,Frerot2018},
by restrictions to the one-particle sector \cite{Gong2014}, or by
using renormalization group techniques \cite{Maghrebi2016}. These
studies suggest nonuniversal behavior, which depends both on the model
but also on the initial condition.

In this work, using a numerically exact method, we study the spreading
of correlations, as measured by $C_{x}\left(t\right)$, for a \emph{generic}
long-range interacting spin-chain. We show that,
\begin{equation}
C_{x}\left(t\right)\sim C_{x}^{\infty}\left(t\right)+A\frac{t}{x^{\alpha}},\label{eq:main_result}
\end{equation}
where $A$ is a constant and $C_{x}^{\infty}\left(t\right)\equiv\lim_{\alpha\to\infty}C_{x}\left(t\right)$,
which constitutes the main result of our work. It implies that up
to logarithmic corrections, the ``light-cone'' is linear for $\alpha>1$
and scales as $t\sim x^{\alpha}$ for $\alpha<1$. 

\emph{Model and method.\textemdash }We study spreading of information
in the long-range spin\textendash $1/2$ XXZ chain,

\begin{equation}
\hat{H}=\sum_{i=1,j\neq i}^{L}\frac{1}{\left|i-j\right|^{\alpha}}\left(\hat{S}_{i}^{+}\hat{S}_{j}^{-}+\hat{S}_{i}^{-}\hat{S}_{j}^{+}+\Delta\hat{S}_{i}^{z}\hat{S}_{j}^{z}\right),\label{eq:ham}
\end{equation}
where $L$ is the size of the system and we set the anisotropy parameter
$\Delta=2$ to break the conservation of the total spin. The total
$z-$projection of the spin is still conserved, and throughout this
work we work in subsectors with smallest positive magnetization. The
model is nonintegrable for all \emph{finite} $\alpha$, but reduces
to integrable models in the $\alpha\to0$ and $\alpha\to\infty$ limits.
For $\alpha<1,$ the energy becomes superextensive invalidating standard
thermodynamics. While this situation can be remedied with a proper
rescaling of the hopping, since in this work we focus only on dynamical
properties, we do not proceed along this route \cite{Kastner2011,Bachelard2013,VanDenWorm2013,Storch2015,Kastner2017}.

To calculate, $C_{x}\left(t\right)$ in (\ref{eq:otoc}) we use the
normalized Frobenius norm, $\|\hat{O}\|_{F}=\sqrt{\mathcal{N}^{-1}\tr\left(\hat{O}^{\dagger}\hat{O}\right)}$,
where $\mathcal{N}$ is the Hilbert space dimension. We set $\hat{A}_{i}\left(t\right)=\hat{S}_{i}^{z}\left(t\right)$
and $\hat{B}_{i+x}=\hat{S}_{i+x}^{z}$, for which $C_{x}\left(t\right)$
reduces to,

\begin{equation}
C_{x}\left(t\right)=\sqrt{\frac{1}{8}-\frac{1}{\mathcal{N}}\tr\left(\hat{S}_{i}^{z}\left(t\right)\hat{S}_{i+x}^{z}\hat{S}_{i}^{z}\left(t\right)\hat{S}_{i+x}^{z}\right)}.\label{eq:otoc_unitary}
\end{equation}
To maximize the available distances, $x,$ in a system of a finite
size, we set $i=3$, namely a short distance from the left boundary
of the system, and restrict our observations to positive $x$ \cite{Luitz2017}.
We have checked that this choice does not introduce a bias for low
enough thresholds at the considered ranges of the interaction since
reflected signals from the left boundary cannot ``catch up'' with
the front propagating directly to the right (see Fig.~\ref{fig:overview}).
For an efficient calculation of $C_{x}\left(t\right)$, we employ
a numerically exact method based on dynamical typicality \cite{Luitz2017}.
In this approach, the trace over the Hilbert space in Eq. (\ref{eq:otoc_unitary})
is approximated with exponential precision (in $L$) by an expectation
value with respect to a random pure state, $\ket{\psi}$, sampled
from the Haar measure \cite{Levy1939}. This allows to reduce the
problem to the calculation of $\bra{\psi}\hat{S}_{i}^{z}\left(t\right)\hat{S}_{i+x}^{z}\hat{S}_{i}^{z}\left(t\right)\hat{S}_{i+x}^{z}\ket{\psi},$which
can be evaluated efficiently by two independent numerically exact
propagations of $\ket{\psi}$ and $\hat{S}_{i+x}^{z}\ket{\psi}$ \cite{Luitz2017}.
The propagation is performed using a Krylov space technique based
on sparse matrix vector products $\hat{H}\ket{\psi}$. While for long-range
interactions, the Hamiltonian matrix is significantly more dense than
for short-range problems, using a massively parallel implementation
we can reach system sizes up to $L=30$ \cite{Luitz2016c}.

\begin{figure}
\includegraphics{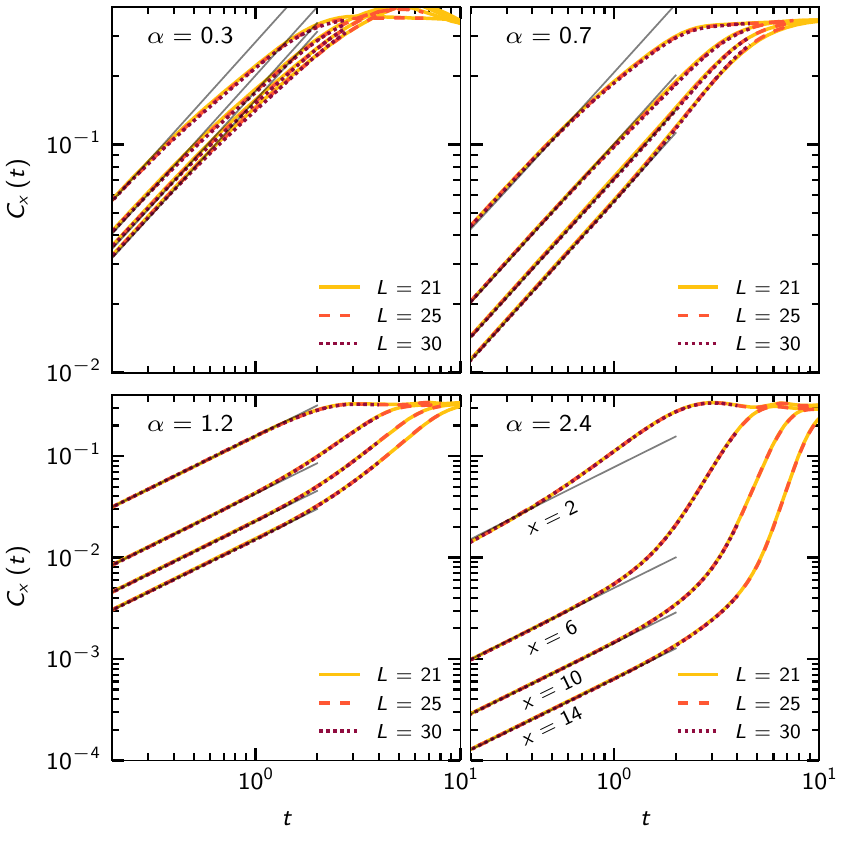}

\caption{\label{fig:temporal}Temporal growth of $C_{x}\left(t\right)$ at
distances $x=2,6,10$, and $14$ (lines order from left to right)
for different interaction exponents $\alpha=0.3,0.7,1.2$ and $2.4$
and system sizes $L=21,25$ and $30$ (indicated by different line
styles). The gray solid lines are linear fits $C_{x}\left(t\right)\sim t$
to the initial temporal growth. }
\end{figure}
\begin{figure}
\includegraphics{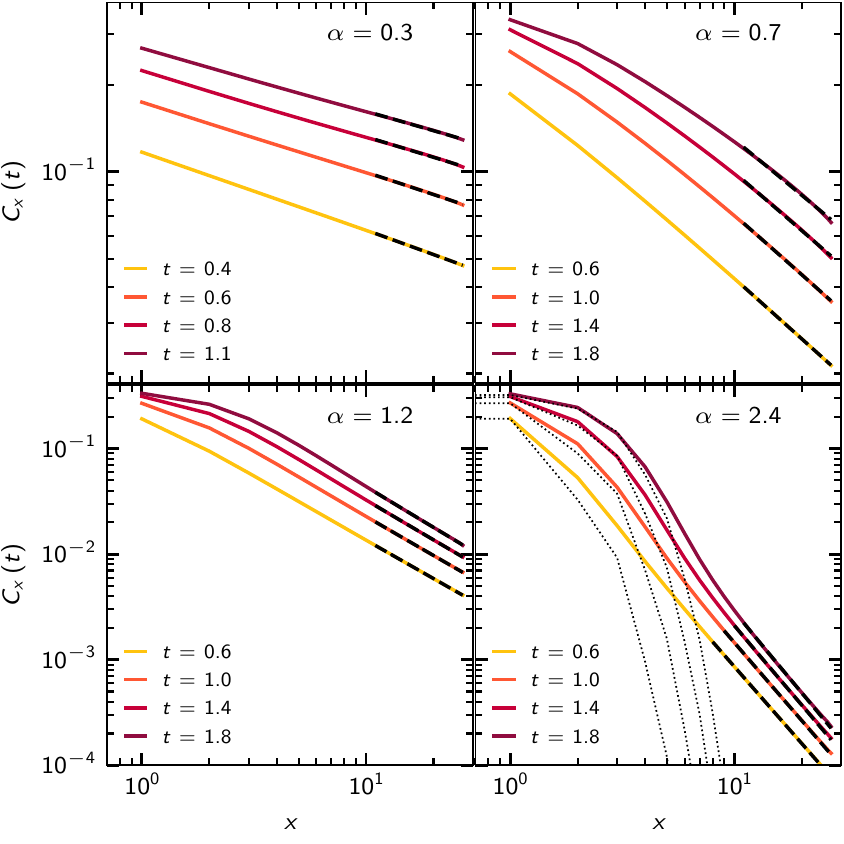}

\caption{\label{fig:spatial}Spatial profiles of $C_{x}\left(t\right)$ for
a few time points $(t\protect\leq2)$ and interaction exponents $\alpha=0.3,0.7,1.2$
and $2.4$. Later time points are indicated by darker colors. Dashed
black lines are power-law fits to the tail of the OTOC. Dotted black
lines correspond to the $\alpha\to\infty$ (short range interaction)
spatial profiles calculated at the same time points as the finite
$\alpha$ data. The system size is $L=30$.}
\end{figure}

\emph{Results.\textemdash }We computed $C_{x}\left(t\right)$ for
various ranges of the power-law interaction, $0<\alpha<3,$ limiting
the propagation to times where $C_{x}\left(t\right)$ saturates in
the entire system. As explained in the introduction, $\alpha=0.5,$$1$
and $2$ (based on analysis of quadratic models) are expected to demarcate
different behavior of $C_{x}\left(t\right)$. We therefore present
representatives $\alpha$-s from each of the four ranges ($\alpha=0.3,0.7,1.2$
and $2.4$). In Fig.~\ref{fig:temporal} we show the initial temporal
growth of $C_{x}\left(t\right)$ at fixed distances $x$ from the
spreading operator, which corresponds to vertical cuts in Fig.~\ref{fig:overview}.
For all values of $\alpha$ we find that the initial temporal growth
is linear in time, and does \emph{not} depend on the distance $x$,
contrary to the situation in quadratic models \cite{Dora2017}. While
for $\alpha<1$, the linear growth is followed by a slower approach
to saturation, for $\alpha>1$, it is followed by a faster than power-law
growth (\emph{cf.} Fig \ref{fig:spatial}). As could be seen from
Fig.~\ref{fig:temporal} the results do \emph{not} depend on the
size of the system in the entire range of $\alpha$ \cite{SuppMat2018}.
 Fig.~\ref{fig:spatial} shows spatial profiles of $C_{x}\left(t\right)$
for a few points of time computed for $L=30$, which corresponds to
horizontal cuts in Fig.~\ref{fig:overview}. We limit our analysis
to times shorter than the saturation regime $\left(t<2\right)$ and
note that for all values of $\alpha$ a regime of power-law decay
of $C_{x}\left(t_{0}\right)\sim x^{-\alpha'}$ is clearly visible.
This regime develops already at very short times and is characterized
by an exponent $\alpha'$, which appears to be time independent. This
implies that for all $\alpha$, information leaking beyond the ``causal''
region is algebraically suppressed . For sufficiently large $\alpha\gtrsim1$,
a visible deformation appears at shorter distances, which corresponds
to the information front of the short-range part of the Hamiltonian.
We demonstrate this by superimposing finite $\alpha=1.6$ and $\alpha=2.4$
spatial profiles with $C_{x}^{\infty}\left(t\right)$, corresponding
to $C_{x}\left(t\right)$ computed for truly short-range interactions
($\alpha\to\infty$), which shows good agreement between the two information
``fronts'' at short distances (the agreement is perfect for larger
$\alpha$). We conclude that the local part of the Hamiltonian is
responsible for the saturation of $C_{x}\left(t\right)$ for $\alpha>1$,
whereas the long-range part is responsible for the power-law tail.
Similar hybrid behavior appears in two of the LR-type bounds \cite{Gong2014,Foss-Feig2015}
and was previously observed for static \cite{Deng2005,Vodola2014,Vodola2015,Gong2016}
and dynamic \emph{one-time} correlation functions \cite{Gong2014,Buyskikh2016},
as also very recently, for two-time correlation functions related
to transport \cite{Kloss2018a}. 

\begin{figure}
\includegraphics{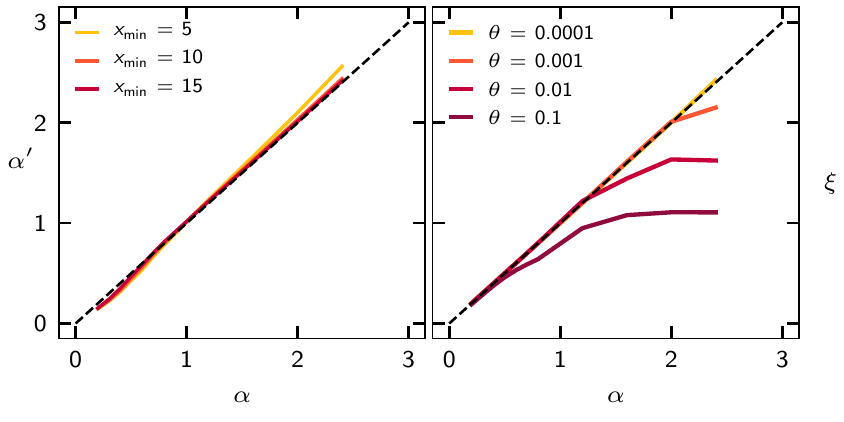}

\caption{\label{fig:exp}\emph{Left}: Exponent $\alpha'$ of the power law
spatial tail of $C_{x}\left(t_{0}\right)$ versus the interaction
exponent $\alpha$ for different fit windows $\left[x_{\text{min}},L-i_{0}\right]$
and fixed time $t_{0}=1$. \emph{Right}: Exponent $\xi$ of the power
law shape of the contour obtained from solving $C_{x}\left(t\right)=\theta$
for different thresholds $\theta$. The dashed lines in both panels
corresponds to $\alpha'=\alpha$ and $\xi=\alpha$ respectively. The
system size is $L=30$.}
\end{figure}
In the left panel of Fig.~\ref{fig:exp} we extracted $\alpha'$
as a function of $\alpha$, by fitting power-laws to $C_{x}\left(t_{0}\right)$,
at a fixed time $t_{0}$. Since the domain of the power-law depends
on $\alpha$, $t_{0}$ and the system size, $L$, we show fit results
for different fit windows $x>x_{\text{min}}$, with $x_{\text{min}}=5,10,15$
to identify the asymptotic behavior at long distances. By increasing
$x_{\text{min}}$ the results appear to converge towards the $\alpha'=\alpha$
line, and do not seem to depend on the system size or the choice of
$t_{0}$ (see Fig.~\ref{fig:spatial}) and \cite{SuppMat2018}. This
spatial dependence of $C_{x}\left(t\right)$ is consistent with all
LR-type bounds for $\alpha>1$, which suggests that the spatial dependence
of all the bounds is tight. Without rescaling of time LR-type bounds
do not hold for $\alpha<1$, and some models show systems size dependence
\cite{Kastner2011,VanDenWorm2013,Storch2015,Kastner2017}. Surprisingly,
even in this regime we find $C_{x}\left(t_{0}\right)\sim x^{-\alpha}$
without notable system size dependence. From Fig.~\ref{fig:temporal}
and Fig.~\ref{fig:spatial} we conclude that for short-enough times
and long-enough distances, $C_{x}\left(t\right)\sim t/x^{\alpha}.$
We confirm this form directly by analyzing the functional dependence
of the contours lines $C_{x}\left(t\right)=\theta$, which for sufficiently
small values of $\theta$ should behave as $t\left(x\right)\propto x^{\xi}$,
with $\xi=\alpha$. The exponent $\xi$ is obtained by numerically
extracting the contours for various values of $\theta$ (see Fig.~\ref{fig:overview})
and fitting it to a power-law behavior for various $\alpha$ (see
right panel of Fig.~\ref{fig:exp}). It is apparent that $\xi$ indeed
converges to $\alpha$ for small thresholds. For $\alpha>2$ power-law
contour lines were rigorously obtained in LR-type bound of Ref.~\cite{Foss-Feig2015},
with an exponent $\xi=1-2/\alpha$. Since we find $\xi=\alpha>1-2/\alpha$,
our results suggest that this bound is not tight and could be improved.

\textit{Discussion.\textemdash }Using a numerically exact technique,
we studied information spreading, as embodied by the out-of-time-order
correlation function, $C_{x}\left(t\right)$ (\ref{eq:otoc}) in a
one-dimensional generic spin-chain with power-law decaying interactions,
$r^{-\alpha}.$ We have shown that for all $\alpha$, sufficiently
far from its saturation value, $C_{x}\left(t\right)\sim t/x^{\alpha}$,
namely it increases linearly in time (see Fig.~\ref{fig:temporal})
and has a power-law decaying tail, with an exponent $\alpha$ (see
Figs.~\ref{fig:spatial} and \ref{fig:exp}). This behavior corresponds
to the leading order in $\left(t/x^{\alpha}\right)$ expansion of
the commutator in Eq.~(\ref{eq:otoc}), indicating that the effect
of the long-range part of the Hamiltonian could be understood perturbatively.
We have confirmed that similar behavior persists for other models
and other local operators taken in (\ref{eq:otoc}), as long as they
are generic (results not shown, but see \cite{SuppMat2018}). Counterintuitively,
the behavior of $C_{x}\left(t\right)$ for $C_{x}\left(t\right)\ll1$
yields sublinear ``light-cones'', $t\sim x^{\alpha}$, with \emph{suppressed}
causal regions for $\alpha>1$. Slower than ballistic behavior, $t\sim x^{\beta\left(\alpha\right)}$
$\left(\beta>1\right)$ was previously observed in the study of spreading
of one-time correlation functions, $K_{x}\left(t\right)$, of quadratic
systems at low-temperatures for $1<\alpha<2$ \cite{Cevolani2015,Cevolani2016,Cevolani2017}.
For $\alpha=3/2$, the exponent $\beta$ was computed analytically
and gives $\beta=\alpha$ \cite{Cevolani2015}, which coincides with
our results for $C_{x}\left(t\right)$. For other values of $\alpha$,
for which $\beta$ was computed numerically, $\beta<\alpha$, but
with a notable upward trend as a function of the systems size \cite{Cevolani2016}.
It is important to note, that while the results obtained from quadratic
models are consistent with ours for short times (or long distances),
it is \emph{not} the case asymptotically, where we find that $C_{x}\left(t\right)$
is well described by the local $\alpha\to\infty$ part of the Hamiltonian
already for $\alpha>1$. 

The overall behavior of $C_{x}\left(t\right)$ we obtain is presented
in Eq.~(\ref{eq:main_result}), and constitutes the main result of
our work. It shows that for $\alpha>1$ the effect of the long-range
part of the Hamiltonian is rather limited, resulting in a transient
behavior were the front which corresponds to the short-range part
is ``catching up'' with the \emph{slower} long-range part. From
Eq.~(\ref{eq:main_result}) and using the LR bound, one can obtain
the asymptotic shape of the ``light-cone,'' which including the
first logarithmic correction is, 
\begin{equation}
t\sim\begin{cases}
\theta x^{\alpha} & \alpha<1\\
x/v-\lambda^{-1}\log\left(\lambda x^{\alpha}\right) & \alpha>1
\end{cases},
\end{equation}
namely faster than ballistic, ``light-cone'' for $\alpha<1$ and
an almost linear, ballistic, ``light-cone'' for $\alpha>1$, with
a finite LR velocity. We note that contrary to quadratic models where
for $1<\alpha<2$ either \emph{subballistic} spreading of correlations
\cite{Cevolani2015,Cevolani2016,Cevolani2017}, or \emph{superballistic}
spreading occurs (depending on the quasiparticle dispersion relation),
we find \emph{ballistic} spreading of correlations, which does not
appear to be model dependent, already for $\alpha>1$ \cite{SuppMat2018}.
This suggests that while low-temperature behavior of generic systems
approximated using quadratic models is highly non-universal, universality
emerges when operator norms are considered. Studying operator norms
therefore allows us to \emph{directly} show that all known LR-type
bounds are \emph{not} tight and could be potentially improved. 
\begin{acknowledgments}
We would like to thank David A. Huse for pointing to us an inconsistency
in the previous version of the manuscript. This project has received
funding from the European Union's Horizon 2020 research and innovation
programme under the Marie Sk\l odowska-Curie grant agreement No. 747914
(QMBDyn). DJL acknowledges PRACE for awarding access to HLRS's Hazel
Hen computer based in Stuttgart, Germany under grant number 2016153659. 
\end{acknowledgments}

\bibliographystyle{apsrev4-1}
\bibliography{lib_yevgeny,local}
\cleardoublepage{}

\subsection{Supplementary Material}

\subsubsection{Contour-lines of the XXZ model}

\begin{figure}[h]
\includegraphics{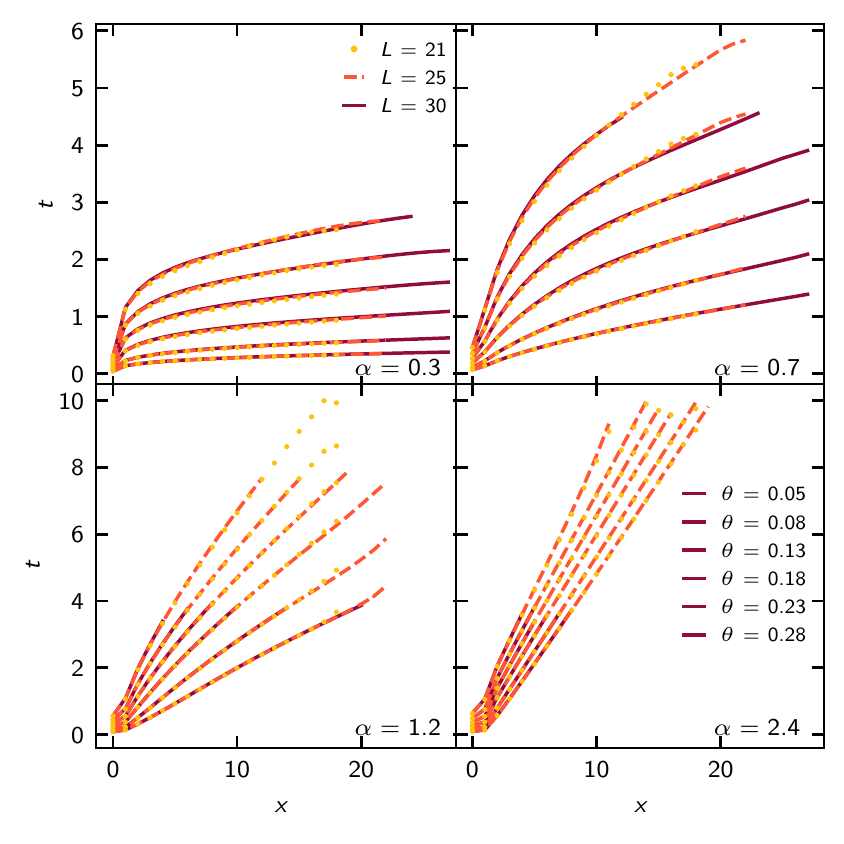}

\caption{\label{fig:contours}Contour lines of the OTOC, obtained from solving
$C_{x}\left(t\right)=\theta$, for six different thresholds $\theta$
(higher $\theta$ corresponds to lines at later times), interaction
exponents $\alpha=0.3,0.7,1.2$ and $2.4$ and $L=21,25$ and $30$
(indicated by different line styles).}
\end{figure}

Since we are interested in bulk effects, it is pertinent to study
the robustness of our results to changes to the size of the system.
Our setup was designed to effectively double the accessible distances
$x$ for the initial excitation, by putting it close to the left boundary.
While reflections from the left boundary become important for $C_{x}\left(t\right)$
close to its saturation value (which is clearly visible in Fig. \ref{fig:overview}),
by focusing on sufficiently small values of $C_{x}\left(t\right)$,
corresponding to the fastest modes, and the information front spreading
to the right $\left(x>0\right)$, such effects could entirely be mitigated.
In particular Fig.~\ref{fig:temporal} in the main text shows that
for small values of $C_{x}\left(t\right)$ the results are independent
of system size for all $\alpha$. For $\alpha<1$ and closer to the
saturation value of $C_{x}\left(t\right)$ a weak system size dependence
is visible. This could also be seen by considering the contour lines
$t_{\theta}(x),$ which correspond to the numerical solution of $C_{x}\left(t\right)=\theta$
for various thresholds $\theta$. In Fig.~\ref{fig:contours} we
show a few $\theta\ll1$ thresholds for three different system sizes.
For $\alpha>1$ the contours for all system sizes agree very well
except for points close to the right boundary of the system, which
appears to introduce a slight artifact. For $\alpha<1$ we observe
a perfect agreement of the contour lines for small thresholds, $\theta$,
whereas larger thresholds lead to discernible finite-size effects.

\subsubsection{Long-range interacting Ising model in a tilted field}

\begin{figure}[h]
\includegraphics[width=1\columnwidth]{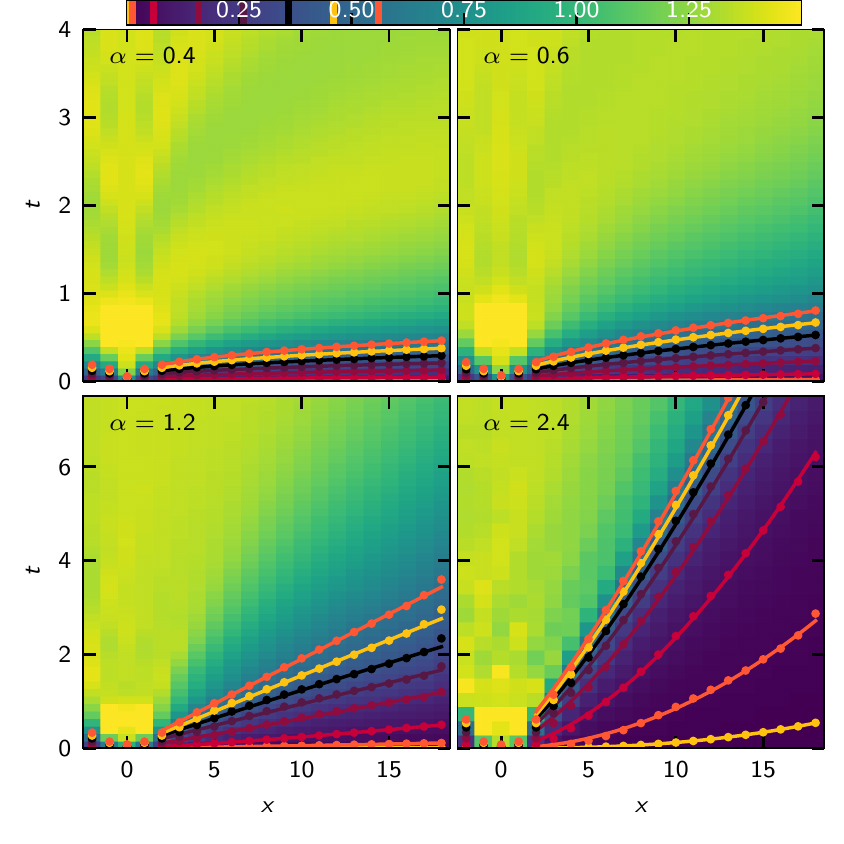}\caption{\label{fig:overview-ising}Spreading of , $C_{x}^{\mathrm{XX}}\left(t\right)$
for various interaction exponents $\alpha$. The points correspond
to discrete contours of the OTOC, calculated by locating where $C_{x}^{XX}\left(t\right)$
exceeds certain thresholds $\theta$, which are indicated in the color
bar. Full lines are power-law fits to these contour lines. Here, we
use open boundary conditions and $L=25$ and 3. }
\end{figure}

In order to check the universality of our findings, we study another
generic long-range model, 

\begin{equation}
\hat{H}_{\mathrm{Ising}}=\sum_{i\neq j}\frac{1}{\left|i-j\right|{}^{\alpha}}\hat{\sigma}_{i}^{z}\hat{\sigma}_{j}^{z}+\sum_{i}\left[h\sin\left(\theta\right)\hat{\sigma}_{i}^{x}+h\cos\left(\theta\right)\hat{\sigma}_{i}^{z}\right],
\end{equation}
which is the long-range interaction Ising model in a tilted magnetic
field (LRTIM). Here $\hat{\sigma}_{i}^{x,z}$ are the Pauli operators
and we will use a field strength of $h=1$ and $\theta=1\text{rad}$,
yielding $h_{x}=0.5403023\dots$ and $h_{z}=0.8414709\dots$ to make
the model generic and nonintegrable in the short range limit $\alpha\to\infty$.
We tested this by confirming that the eigenvalue statistics correspond
to GOE (results not shown).

While in the main text, we considered the OTOC, $C_{x}\left(t\right)$,
defined by the commutator of two initially local $\hat{S}_{i}^{z}$
operators, this commutator turns to be special for the Ising model,
since the first two leading terms in the expansion of the commutator,
$\left[\hat{S}_{i}^{z}\left(t\right),\hat{S}_{j}^{z}\right]$ with
respect to time vanish. This confirms our statement in the main text
that the short-time behavior seems to be captured by perturbation
theory, and highlights the importance of the choice of generic operators
for the calculation of the OTOC. We therefore consider the OTOC defined
by $\hat{\sigma}^{x}$ operators, for which the leading term in perturbation
theory does \emph{not} vanish,

\begin{equation}
C_{x}^{\mathrm{XX}}\left(t\right)=\left\Vert \left[\hat{\sigma}_{i}^{x}\left(t\right),\hat{\sigma}_{i+x}^{x}\right]\right\Vert .
\end{equation}
Similarly to the main text we will fix the operator norm to the Frobenius
norm. We note in passing that we do not rescale the parameters of
the Hamiltonian with system-size dependent factors and therefore the
model is pathological for $\alpha<1$, where the long-range part becomes
superextensive and therefore dominant in the thermodynamic limit.
As we point out in the main text, the behavior for $\alpha<1$ is
not universal and we therefore do not consider this limit. 

In Fig. \ref{fig:overview-ising}, we show our result of the OTOC,
$C_{x}^{\mathrm{XX}}\left(t\right)$, for the LRTIM for different
interaction exponents $\alpha$. As for the XXZ model, we observe
power law shapes of the contour lines of the OTOC for various thresholds
$\theta$.

\begin{figure}[h]
\includegraphics{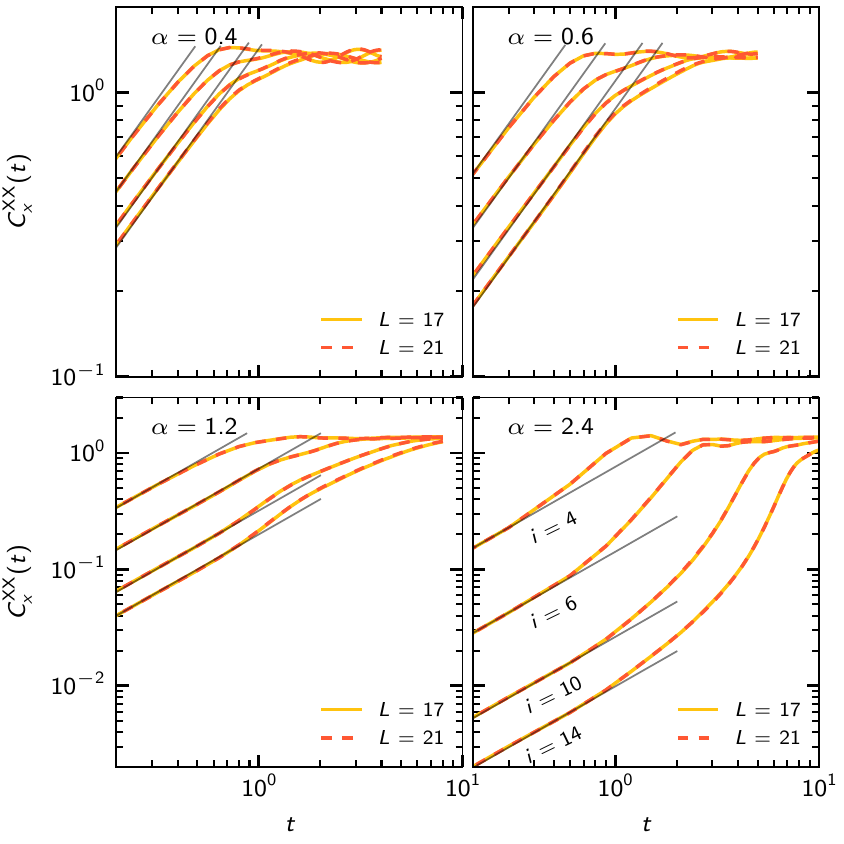}

\caption{\label{fig:temporal-ising}Temporal growth of $C_{x}^{XX}\left(t\right)$
at distances $x=4,6,10$, and $14$ (lines order from left to right)
for different interaction exponents $\alpha=0.4,0.6,1.2$ and $2.4$
and system sizes $L=17$ and $21$ (indicated by different line styles).
The gray solid lines are linear fits $C_{x}^{\mathrm{XX}}\left(t\right)\sim t$
to the initial temporal growth. }
\end{figure}

To scrutinize our finding of the linear growth of the OTOC for short
times, independently of the value of $\alpha$, we fit linear functions
(with zero intercept) $C_{x}^{\mathrm{XX}}\left(t\right)=\theta$
to the early time growth of $C_{x}^{\mathrm{XX}}\left(t\right)$ in
the LRTIM, which perfectly capture the initial growth regime. Interestingly,
even though in the LRTIM the long-range part overwhelms the local
field term for large system sizes and $\alpha<1$, we do not observe
visible finite size effects at early times.

\begin{figure}[h]
\includegraphics{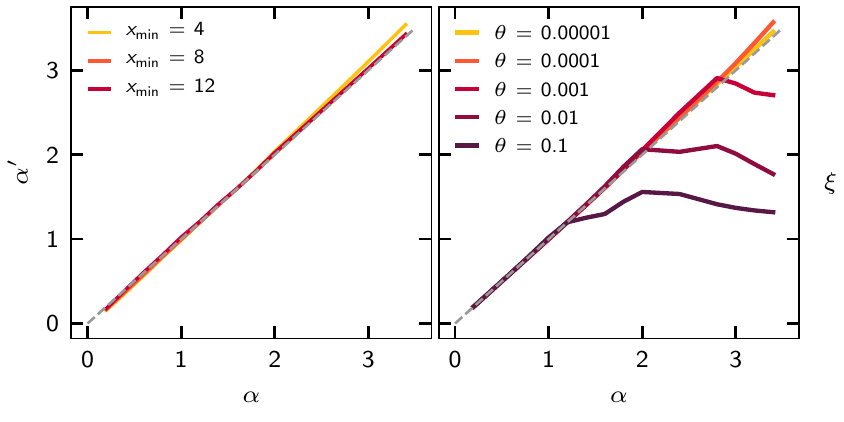}

\caption{\label{fig:exp-ising}\emph{Left}: Exponent $\alpha'$ of the power
law spatial tail of $C_{x}^{\mathrm{XX}}\left(t_{0}\right)$ versus
the interaction exponent $\alpha$ for different fit windows $\left[x_{\text{min}},L-i\right]$
and fixed time $t_{0}=1$. \emph{Right}: Exponent $\xi$ of the power
law shape of the contour obtained from solving $C_{x}^{\mathrm{XX}}\left(t\right)=\theta$
for different thresholds $\theta$. The dashed lines in both panels
corresponds to $\alpha'=\alpha$ and $\xi=\alpha$ respectively. The
system size here is $L=21$.}
\end{figure}

\begin{figure}
\includegraphics{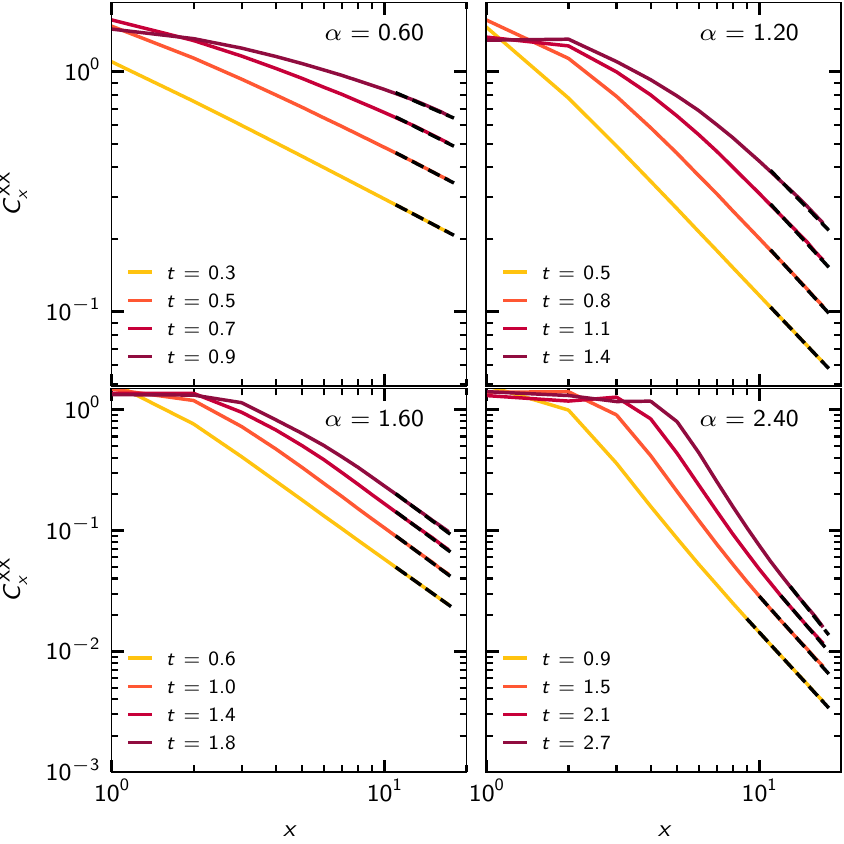}

\caption{\label{fig:spatial-ising}Spatial profiles of $C_{x}^{\mathrm{XX}}\left(t\right)$
in the long-range Ising model for a few time points $(t\protect\leq2)$
and interaction exponents $\alpha=0.6,1.2,1.6$ and $2.4$. Later
time points are indicated by darker colors. Dashed black lines are
power-law fits to the tail of the OTOC. The system size is $L=21$.}
\end{figure}

The analysis of the OTOC $C_{x}^{\mathrm{XX}}\left(t\right)$ for
the LRTIM provides additional support to the asymptotic form of the
``light-cone,'' obtained in the main text for the long-range XXZ
model. We analyze the long distance tails of the OTOC at different
fixed times in Fig.~\ref{fig:spatial-ising} for various values of
the interaction exponent $\alpha$ and fit power law tails of the
form $x^{\alpha'}$. The corresponding exponent $\alpha'$ is shown
in the left panel of Fig.~\ref{fig:exp-ising}, confirming that also
in the LRTIM for long enough distances the exponent $\alpha'$ converges
to the same value as the interaction exponent $\alpha$.

In the right panel of Fig. \ref{fig:exp-ising}, we show the exponent
of power law fits to the contours obtained from the solution of $C_{x}^{\mathrm{XX}}\left(t\right)=\theta$
for different thresholds $\theta$ as a function of the interaction
exponent $\alpha$. Some of these contours and their fits are displayed
in Fig.~\ref{fig:overview-ising}. This analysis confirms our finding
that for small enough thresholds, the contour exponent matches the
interaction exponent: $\xi=\alpha$, capturing the ``fast'' modes
of information propagation which are only due to the long range nature
of interactions. For larger thresholds and $\alpha>1$, this is not
the case, because the information ``front'' due to the short range
part of the Hamiltonian, which is \emph{not} of power law form, catches
up.

In summary, our analysis of another generic spin-chain with long-range
interactions confirms all our findings presented in the main text
and demonstrates their universality.
\end{document}